\documentclass[fleqn,usenatbib,useAMS]{mnras}
\usepackage{graphicx}   
\usepackage{amsmath}    
\usepackage{amssymb}    
\usepackage{multicol}        
\usepackage{bm}         
\usepackage{pdflscape}  
\usepackage[T1]{fontenc}
\usepackage{ae,aecompl}
\usepackage{newtxtext,newtxmath}

\title[G183$-$35]{A Magnetic White Dwarf with Five H$\alpha$ Components}
\author[Kilic et al.]
{Mukremin Kilic$^1$,
B. Rolland$^2$,
P. Bergeron$^2$,
Z. Vanderbosch$^{3,4}$,
P. Benni$^5$,
J. Garlitz$^6$
\\
$^1$Homer L. Dodge Department of Physics and Astronomy, University of Oklahoma, 440 W. Brooks St., Norman, OK, 73019, USA\\
$^2$D\'epartement de Physique, Universit\'e de Montr\'eal, C.P. 6128, Succ. Centre-Ville, Montr\'eal, QC H3C 3J7, Canada\\
$^3$Department of Astronomy, University of Texas at Austin, Austin, TX, 78712, USA\\
$^4$McDonald Observatory, Fort Davis, TX, 79734, USA\\
$^5$Acton Sky Portal, 3 Concetta Circle, Acton, MA 01720, USA\\
$^6$Private Observatory, 1155 Hartford St, Elgin, OR 97827, USA\\
}

\date{\ \ Submitted \today \vspace{-0.5cm}}
\pubyear{2019}

\begin{document}
\label{firstpage}
\pagerange{\pageref{firstpage}--\pageref{lastpage}}
\maketitle

\begin{abstract}
G183$-$35 is an unusual white dwarf that shows an H$\alpha$ line split into five components, instead of the usual
three components seen in strongly magnetic white dwarfs. Potential explanations for the unusual set of lines includes a double
degenerate system containing two magnetic white dwarfs and/or rotational modulation of a complex magnetic field structure. Here
we present time-resolved spectroscopy of G183$-$35 obtained at the Gemini Observatory. These data reveal two sets of
absorption lines that appear and disappear over a period of about 4 hours. We also detect low-level (0.2\%) variability
in optical photometry at the same period. We demonstrate that the spectroscopic and photometric variability can be explained
by the presence of spots on the surface of the white dwarf and a change in the average field strength from about 4.6 MG to
6.2 MG. The observed variability is clearly due to G183$-$35's relatively short spin period. However, rotational modulation of a complex
magnetic field by itself cannot explain the changes seen in the central H$\alpha$ component. An additional source of variability in
the line profiles, most likely due to a chemically inhomogeneous surface composition, is also needed. We propose further observations
of similar objects to test this scenario.
\end{abstract}


\section{Introduction}

About 10 to 20\% of white dwarfs in the solar neighborhood are strongly magnetic, with field strengths of up to $10^9$ G
\citep{kawka07,brinkworth13,ferrario15}. Weak fields, $B\leq$ 1 kG, may also be present in most white dwarfs,
but are hard to detect  \citep{jordan07,landstreet12}. Magnetic white dwarfs tend to be on average
higher in mass compared to their non-magnetic counterparts \citep{liebert88,kawka07,briggs15}. Curiously, high field magnetic white dwarfs are never
found in wide binary systems with late-type stellar companions \citep{liebert05}. This led to \citet{tout08} and \citet{briggs15,briggs18}
to suggest that merging binaries within a common envelope can explain the incidence of magnetism and the mass
distribution of high-field magnetic white dwarfs.

Several high field magnetic white dwarfs are confirmed to be in common proper motion or short period binaries with other white dwarfs
\citep{ferrario97,girven10,dobbie12}.  For example, NLTT 12758 is a magnetic white dwarf with a non-magnetic
companion white dwarf in a 1.154 d orbit \citep{kawka17}. \citet{rolland15} analyzed 16 magnetic DA white dwarfs with
high signal-to-noise ratio optical spectroscopy available, and found that offset dipole models can explain six of these stars.
However, the remaining 10 stars in their sample have photometric temperatures that are inconsistent with their spectroscopy,
and these may be in unresolved binary systems. However, the overabundance of binary candidates in such a small sample of high-field
magnetic white dwarfs is intriguing. One of these stars, G183$-$35 (also known as NLTT 46206 and WD 1814+248)
displays an H$\alpha$ line that is split into five components, which could be due to a combination of two magnetic DA white dwarfs
in this system \citep{rolland15}.
 
G183$-$35 was identified as a high proper motion object by \citet{giclas71} and classified to
be a DC white dwarf by \citet{hintzen74} based on low-resolution spectroscopy. \citet{putney95} performed a spectropolarimetric
survey of several white dwarfs, including G183$-$35, and detected H lines split into three components 
due to a magnetic field strength of  $6.8 \pm 0.5$ MG.  In addition, she found evidence of a change in both line shapes and
polarization spectra taken more than a year apart, and interpreted this as evidence of rotation in this object. To search for
a rotation period,  \citet{brinkworth13} obtained follow-up photometry of G183$-$35, but did not find any evidence
of variability at  the $\geq$4\% level on timescales of less than a year. 

To explore the origin of the unusual splitting of the H$\alpha$ line in G183$-$35, we obtained time-resolved spectroscopy
and photometry over multiple nights. Here we present the results of this study. We list the details of our observations in section 2,
discuss the variability in the line shapes, radial velocity of the central H$\alpha$ line, and photometry in section 3.
We constrain the physical parameters of G183$-$35, including its rotation period, in section 4, and conclude in section 5.
 
\section{Observations}

\begin{figure}
\includegraphics[width=3.5in]{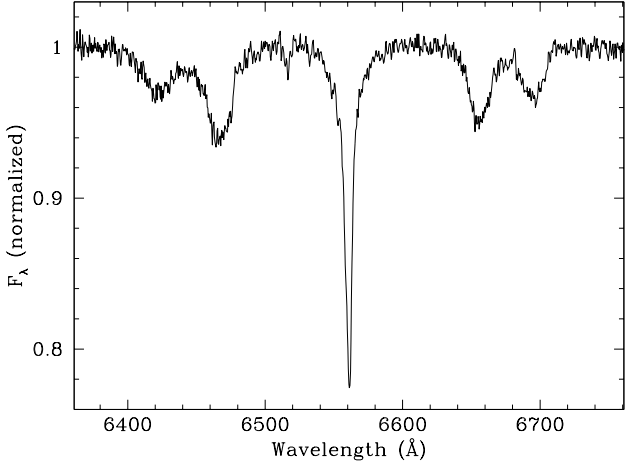}
\vspace{-0.1in}
\caption{The average spectrum of G183$-$35 based on our Gemini data. This spectrum has a signal-to-noise ratio of 250 in
the continuum, and reveals a central H$\alpha$ line along with two other sets of Zeeman-split lines that are separated from the
central component by about 95 \AA\ and 140 \AA, respectively. Note that the weak absorption feature near 6516 \AA\ is telluric.}
\label{fig:gem}
\end{figure}

We obtained follow-up optical spectroscopy of G183$-$35 using the 8m Gemini North telescope equipped 
with the Gemini Multi-Object Spectrograph (GMOS) as part of the Fast Turnaround queue program GN-2017B-FT-2.  We obtained
a sequence of 32 $\times$ 5 min long back-to-back exposures on UT 2017 Sep 10 with the R831 grating and a
0.5$\arcsec$ slit, providing wavelength coverage from 5350 \AA\ to 7710 \AA\ and a resolution of 0.376 \AA\ per pixel.
Each spectrum has a comparison lamp exposure taken within 10 min of the observation time. We obtained additional
sets of 4 and 10 back-to-back exposures on UT 2017 Sep 11 and 12, respectively. We used the IRAF Gemini gmos package
to reduce these data. Figure \ref{fig:gem} shows our summed Gemini spectrum based on 46 exposures. This spectrum
has a signal-to-noise ratio of 250 in the continuum, and clearly shows a narrow central H$\alpha$ line, and four other Zeeman-split
H$\alpha$ lines as noted by \citet{rolland15}.

We obtained follow-up V-band optical photometry of G183$-$35  with 1 min long exposures over 5.9 h on
UT 2017 June 15 using a 35-cm Schmidt-Cassegrain telescope at the Acton Sky Portal in Massachusetts.
We also obtained white-light optical photometry of the same target with 6 min long exposures over 5.5 h on
UT 2017 June 22 using a Celestron 28-cm telescope at a private observatory in Oregon. 

We obtained additional follow-up time-series photometry on UT 2017 June 21-25 and July 26-29 using the
McDonald Observatory 2.1m Otto Struve telescope with the ProEM camera and the BG40 filter. We used
exposure times of 10 to 30 s with a total integration time of 19.88 h. We binned the CCD by $4\times4$, which resulted in a
plate scale of $0.38\arcsec$ pixel$^{-1}$. We used several comparison stars to correct for
transparency variations.

\section{Results}

\subsection{Spectroscopic Variability}

\begin{figure}
\includegraphics[width=3.6in]{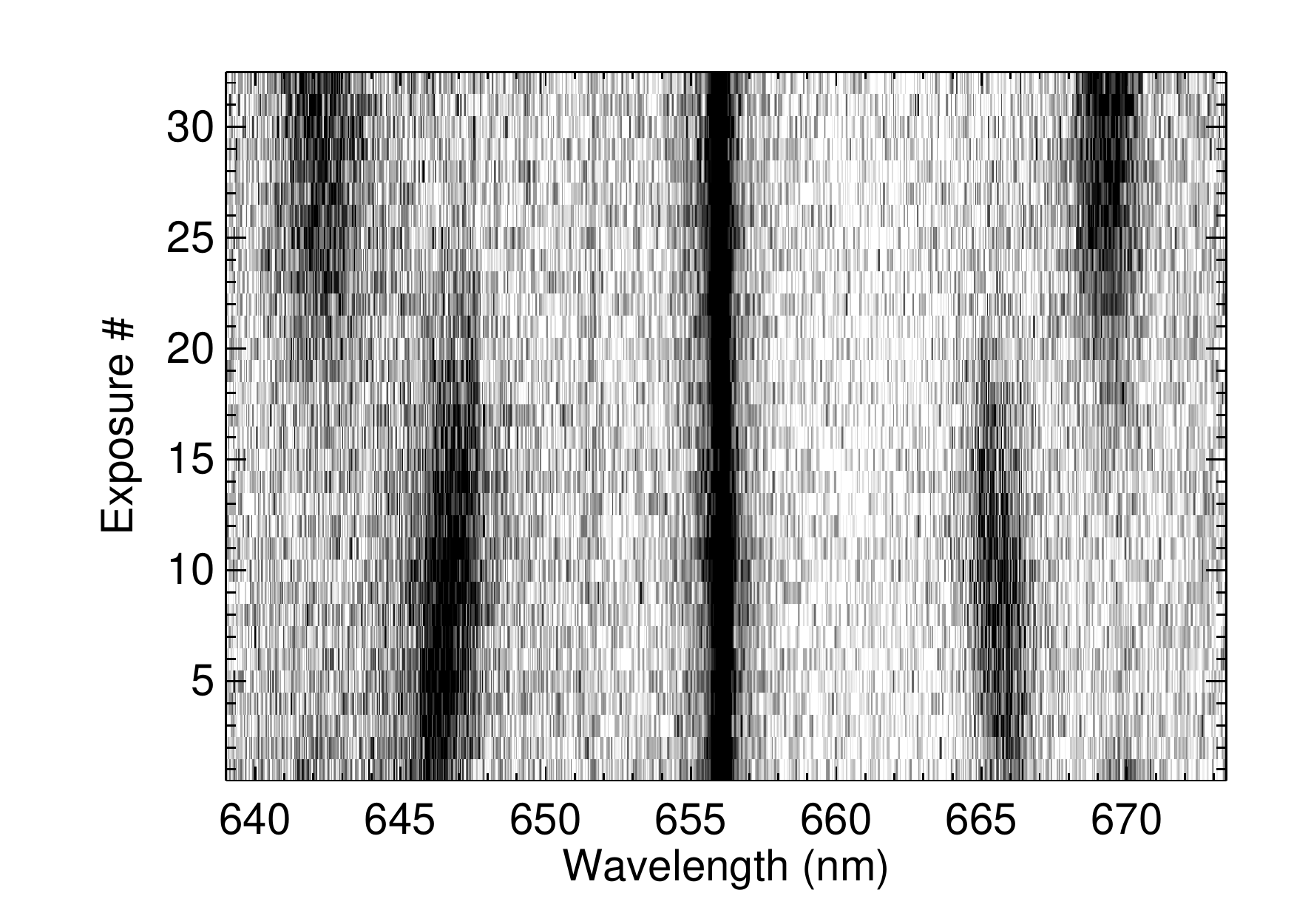}
\includegraphics[width=3.6in]{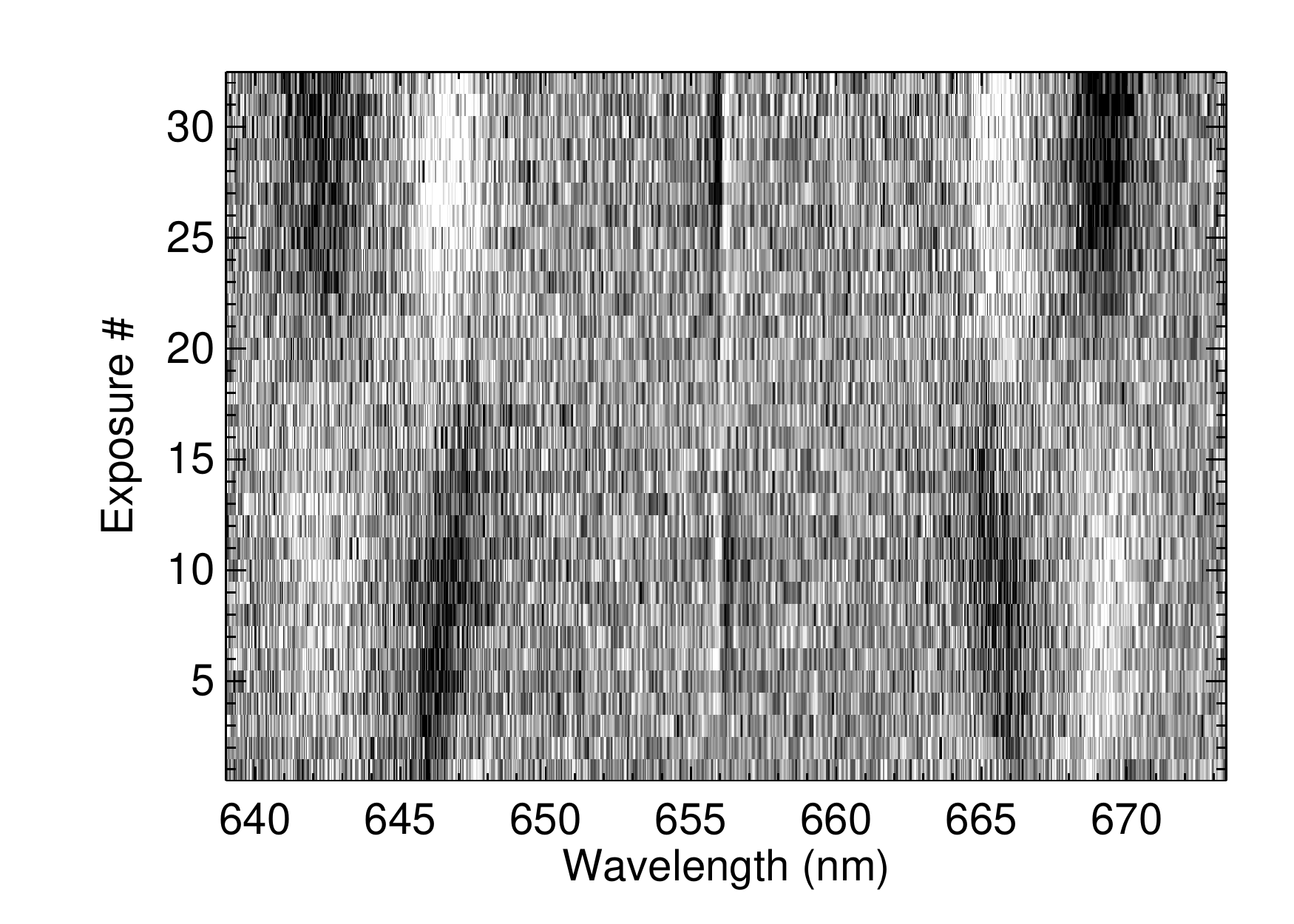}
\caption{{\it Top:} Gemini time-resolved spectroscopy of G183$-$35 over 2.9 hours on UT 2017 Sep 10.
{\it Bottom:} The difference image between each spectrum shown above and the average spectrum.}
\label{fig:trail}
\end{figure}

Figure \ref{fig:trail} shows the Gemini/GMOS trailed spectrum of G183$-$35 based on 32 back-to-back exposures taken
over 2.9 hours on UT 2017 Sep 10. This figure reveals relatively quick changes in the line profiles. The central H$\alpha$
component is always there, but the other components, the inner and outer sets of lines appear and disappear in sequence. 
The first exposure has both sets of lines visible (the inner and outer pairs), but then only the inner pair is visible with
a separation of $\approx100$ \AA\ from the central component. The inner pair stays visible for about 20 exposures (each exposure
is 5 min long), but its separation from the central component decreases over time to $\approx$90 \AA. At this point, the inner
pair of lines disappear, and the wider pair of lines become visible. The wider pair remains at about the same wavelength for the rest of the
observations on the same night.

We have an additional set of 14 spectra taken on consecutive nights (not shown here). These spectra also display
the change in the inner pair of lines over short timescales. We measured the equivalent widths of the inner and outer
pair of lines in each spectrum, and found three significant peaks at 0.142, 0.153, and 0.166 d in a Lomb-Scargle
diagram \citep{lomb76,scargle82}  of these equivalent width measurements. 

The bottom panel in Figure \ref{fig:trail} shows the differences between the trailed spectrum of G183$-$35 (shown in the top panel)
and its average spectrum. The shift in wavelength of the inner pair of lines compared to the average spectrum is clearly visible. The
missing absorption features in each spectrum (compared to the average) appear brighter. More importantly,
this figure also reveals a significant shift in the wavelength of the central H$\alpha$ line over time. To verify that this shift in
the central component is not due to the spectrograph flexure, we used the telluric
lines between 7160 and 7400 \AA\ in each Gemini spectrum. We measured an average velocity offset
of $-0.2 \pm 2.5$ km s$^{-1}$ in this wavelength range. Hence, the systematic errors in our radial velocity
measurements are on the order of only a few km s$^{-1}$.

\begin{figure}
\includegraphics[width=3.4in]{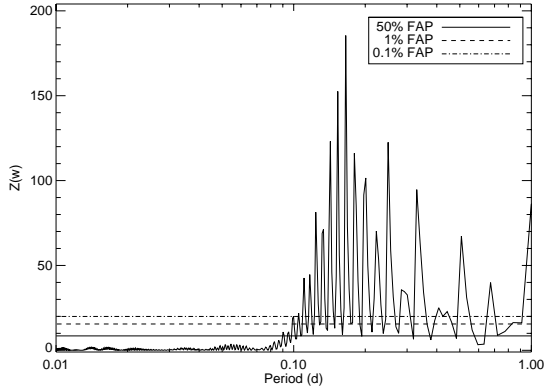}
\includegraphics[width=3.4in]{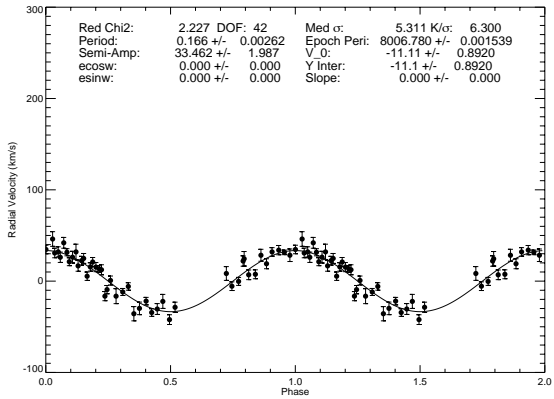}
\caption{{\it Top:} Lomb-Scargle periodogram of the radial velocity of the central H$\alpha$ component. The solid,
dashed, and dashed-dotted lines mark the 50\%, 1\%, and 0.1\% False-Alarm Probability (FAP) limits.
{\it Bottom:} The best-fitting solution assuming a circular orbit.}
\label{fig:rv}
\end{figure}

We used the cross-correlation package RVSAO \citep{kurtz98} to measure the radial velocity of the H$\alpha$ line in the wavelength
range $6500 - 6620$ \AA. We used the average spectrum of G183$-$35 as the template spectrum, since
we are only interested in constraining the relative shifts in the central H$\alpha$ line. Our final velocities come
from cross-correlating the individual observations with this template. 

To search for periodicities in the radial velocity data, we computed the Lomb-Scargle periodogram using the IDL
program MPRVFIT \citep{delee13}. Figure \ref{fig:rv} shows the Lomb-Scargle periodogram and the best-fitting solution
assuming a circular orbit. The highest peak is at $P= 0.166$ d, which implies a velocity semi-amplitude of
$K= 33.5 \pm 2.0$ km s$^{-1}$ and a mass function of $f=0.00064 \pm 0.00011 M_{\odot}$. 
However, just like in the equivalent width measurements of the inner and outer pair of lines discussed above, there are
several significant aliases in the periodogram, including 0.142 and 0.153 d. Note that these frequencies
are separated from each other by the daily alias (11.57 $\mu$Hz), which makes it hard to identify the exact frequency
of variation. 

\subsection{Photometric Variability}

\begin{figure}
\includegraphics[width=3.6in]{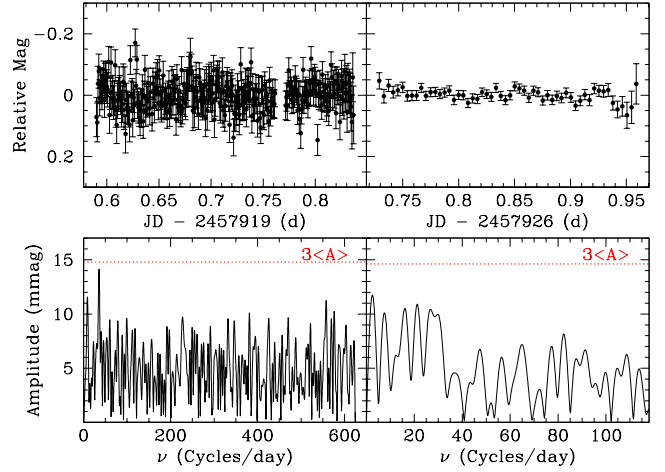}
\vspace{-0.1in}
\caption{Time series photometry of G183$-$35 obtained over 5.9 h and 5.5 h at
Acton Sky Portal (top left) and a private observatory in Oregon (top right) on two different nights in 2017 June.
The bottom panels show the Fourier Transform and the 3 $<A>$ detection limit for each dataset, where
$<A>$ is the average amplitude in the frequency range shown.}
\label{fig:amat}
\end{figure}

\begin{figure*}
\hspace{-0.2in}\includegraphics[width=7in]{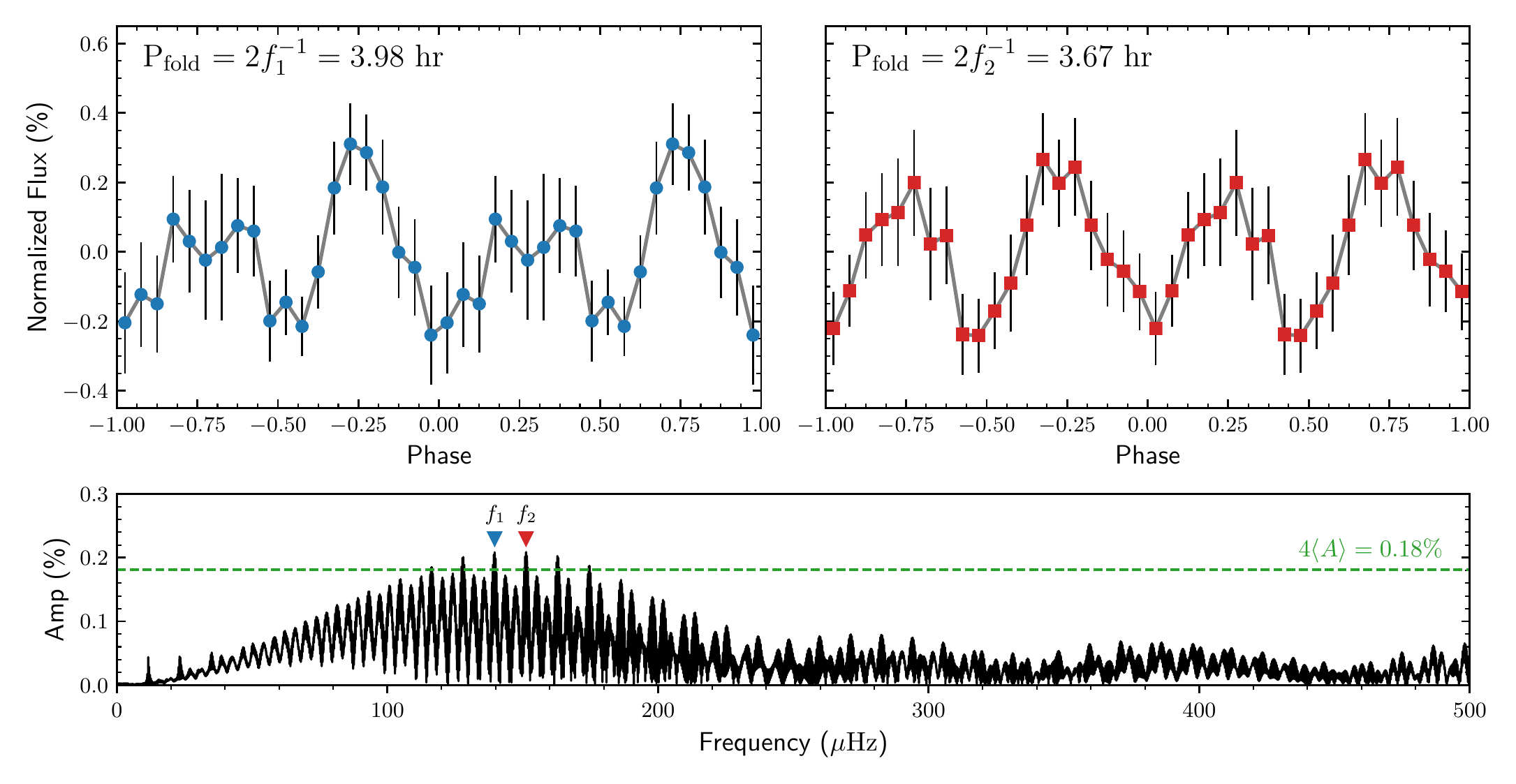}
\caption{Time-series photometry of G183$-$35 obtained over 19.88 h at the McDonald Observatory 2.1m telescope.
The data is binned into 20 equally spaced bins per full phase, folded at 3.98 h (top left) and 3.67 h (top right).
The bottom panel shows the periodogram for all of the McDonald data combined. The two highest peaks in
the periodogram are labelled, and the dotted line shows the 4$<A>$ detection limit.}
\label{fig:mcd}
\end{figure*}

\citet{brinkworth13} observed G183$-$35 over a week in 2002 August and another week in 2003 May, and found no
evidence of variability or rotation on timescales of less than a year. They concluded that this star is not varying at the 4\%
peak-to-peak level on these timescales. Figure \ref{fig:amat} shows time series photometry of G183$-$35 obtained
over two different nights in 2017 June. These data come from the 28-35 cm telescopes in Massachusetts and Oregon,
and slightly improve the limits on variability to a 3$<A>$ detection limit of 15 mmag.  

Figure \ref{fig:mcd} shows the time-series photometry of the same target obtained at the McDonald Observatory
2.1m telescope. The bottom panel shows the periodogram of all of the McDonald data combined. These data
improve the limits on variability significantly, to a $4<A>$ limit of only 0.18\%. The periodogram for the combined data
shows lots of aliasing, but the highest peak is at 151.2 $\mu$Hz (period of 1.83 hours) with an amplitude of 0.21\%.
If we treat the June and July data separately, the highest peak shifts to 127.9 $\mu$Hz with amplitude 0.26\% and 
174.1 $\mu$Hz with amplitude 0.22\%, respectively. 

Considering that these frequencies are all offset from each other by integer multiples of the daily alias (11.57 $\mu$Hz),
these data sets are all consistent with the same frequency. The period that we get from the Gemini radial velocity data
(3.98 hours) has a half period of 1.99 hours and a corresponding frequency of 139.6 $\mu$Hz. This is one daily alias
away from the highest peak in both the June and the combined data, and three daily aliases away from the peak in the July
data. Hence, even though the observed photometric variability is small, the detection is significant, and it matches
the observed variability in the radial velocities of the central H$\alpha$ component, and also the changes in the equivalent
width measurements of the inner and outer pair of lines. 

The top panels in Figure \ref{fig:mcd} show the phase-folded McDonald light curve using the best-fit period from the spectroscopy
(3.98 h) or photometry data (3.67 h). Both are acceptable due to the aliasing present, and both light curves show that G183$-$35
shows $\approx0.4$\% peak-to-peak variations over about 4 h.

\section{Discussion}

\subsection{Is This a Binary System?}

\citet{leggett18} used the U.S. Naval Observatory parallaxes and photometry to derive 
$T_{\rm eff} = 6870 \pm 170$ K, $\log{g}= 8.07 \pm 0.06$, and $M = 0.63 \pm 0.05 M_{\odot}$ for G183$-$35
under the assumption of a single star. However, they also noted that there is a significant
discrepancy between the USNO and Gaia Data Release 2 \citep{gaia18} parallax measurements for four of the stars in their sample, including G183$-$35. 
One of these stars (WD 0239+109) is a known double degenerate system, and \citet{leggett18} suggest that perhaps all
four of these stars with discrepant USNO parallaxes are binary systems. Using the Gaia parallax, the best-fit parameters for
G183$-$35 under the assumption of a single star change to $T_{\rm eff} =  6770 \pm 200 $ K, $\log{g}=8.28 \pm 0.05$,
and $M = 0.77 \pm 0.04 M_{\odot}$. 

A single magnetic DA white dwarf model
with a temperature near 6800 K overpredicts the central component of the observed H$\alpha$ line profile in G183$-$35.
To match the central component, the temperature needs to be reduced to about 5600 K, which is clearly inconsistent with the
observed photometry. We note that there is another Gaia source (4578913734331945984) with $G=19.89$ mag,
$G_{BP}-G_{RP}=1.33$ mag, and within $3.4\arcsec$ of G183$-$35. This other star is likely the source of the
observed mid-infrared excess in the WISE photometry of this object \citep{leggett18}, but it is unlikely to affect the
USNO parallax measurements. 

\citet{rolland15} performed a photometric and spectroscopic deconvolution of the suspected unresolved binaries in their sample
by diluting their magnetic DAH white dwarf models with DC, DA, or DAH companions. They fit the temperatures and radii of both
components, and identified 8, 1, and 1 systems with likely DC, DA, and DAH white dwarf companions, respectively. They obtained
temperatures of 5998 and 5849 K and $R_B / R_A = 1.128$ for the two magnetic white dwarf candidates in G183$-$35.  
These temperature and radii ratios along with the Gaia parallax imply a binary system containing
a $0.97M_{\odot}$ primary and a $0.87M_{\odot}$ secondary. 

If this is a double white dwarf system, and if the observed radial velocity variations of the central Halpha component (with
semi-amplitude 33.5 km/s) are due to orbital motion, this implies a minimum mass companion of $0.09M_{\odot}$ for a $0.97M_{\odot}$ primary. Hence,
if the velocity changes are due to orbital motion of a $0.97 + 0.87 M_{\odot}$ binary, this would require a low-inclination ($i\approx9^{\circ}$) system. However, orbital
motion in such a low inclination (almost face on) system cannot explain the inner and outer pair of the H$\alpha$ components
appearing and disappearing over several hours. An almost equal-mass binary would show significantly larger radial velocity shifts. For example the $P=1.154$ d binary NLTT12758
consists of an $M=0.69 M_{\odot}$ magnetic white dwarf with an $M=0.83 M_{\odot}$ non-magnetic companion and displays
up to 200 km/s radial velocity variations. Interestingly NLTT 12758 shows photometric variations every 23 minutes due to the
fast spinning magnetic white dwarf in that system. Hence, the observed spectroscopic variability in G183$-$35
is not due to orbital motion, but rather from changes in the line profiles due to the rotation of the magnetic white dwarf.

\subsection{Rotational Modulation}

Photometric and spectroscopic variations due to rotation are commonly
observed in magnetic white dwarfs.  Photometric variability can be due
to star spots or magnetic dichroism in high-field white dwarfs
\citep{ferrario97}, whereas spectroscopic variability is usually
caused by variations in the surface field strength that impacts the
Zeeman-split components \citep{brinkworth13}.

To study the variability of the magnetic field structure on G183$-$35,
we use offset dipole models to fit each Gemini spectrum under the
assumption of a single magnetic DA white dwarf. The offset dipole
models treat the dipole field strength $B_d$, viewing angle $i$, and
the offset $a_z$ (in units of stellar radii) as free parameters to
reproduce the observed spectrum, and thus the magnetic field
distribution across the stellar surface. Figure 4 of
\citet{bergeron92} illustrates the flexibility of the offset dipole
models to match the field distribution across the stellar surface. In
addition to $B_d$, $i$, and $a_z$, we also treat the effective
temperature as a free parameter to get a reasonable fit to both the
central H$\alpha$ component and the inner/outer pair of lines. The
viewing angle changes the asymmetry of the shifted components of the
H$\alpha$ line. We found that a viewing angle of $\sim30^{\circ}$
gives the best match to the data, and therefore we kept it constant in
our fits.

\begin{figure}
\hspace{-0.3in}
\includegraphics[width=4.0in]{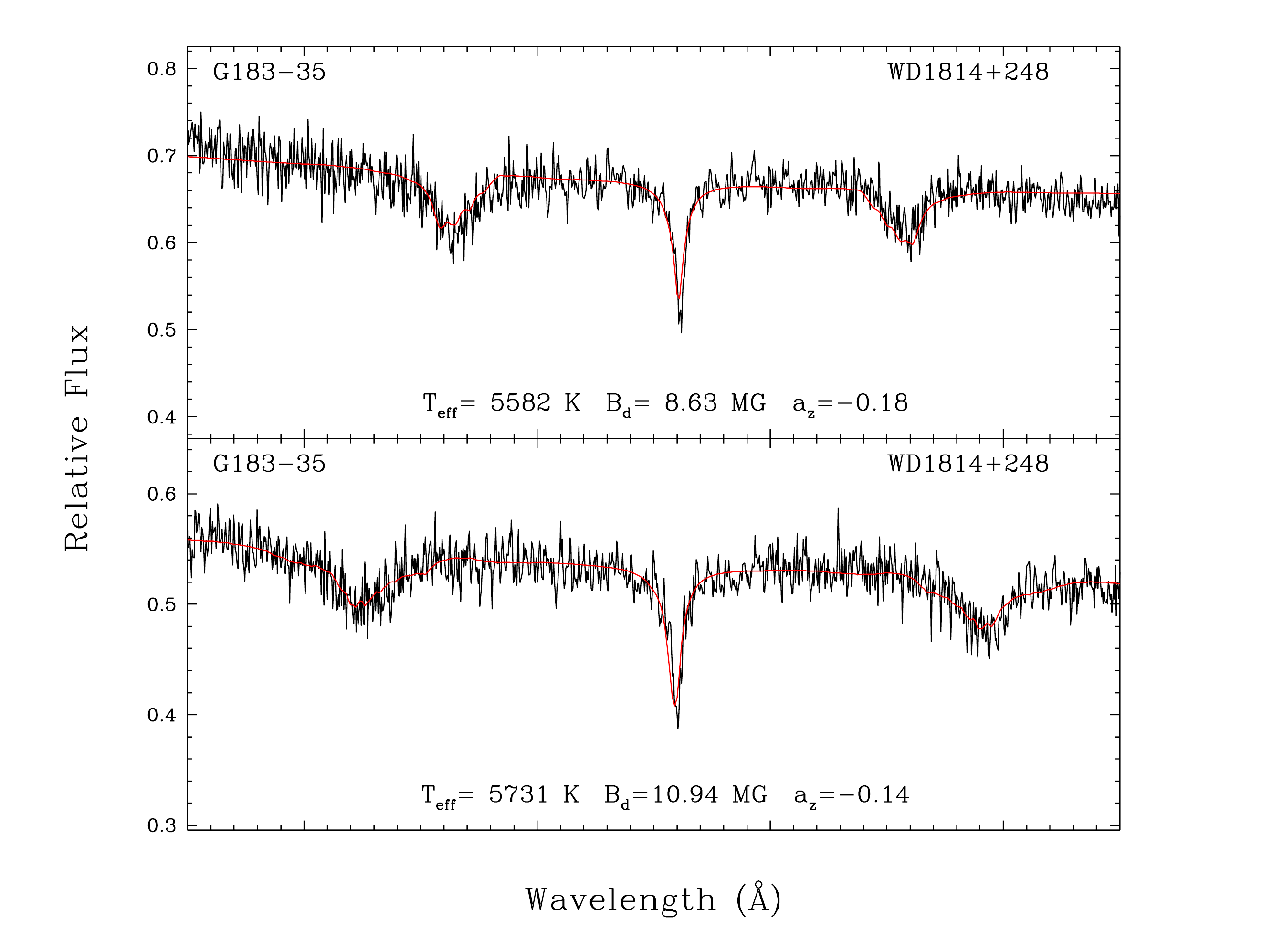}
\vspace{-0.2in}
\caption{Our best fits (red lines) to two of the Gemini spectra using
  offset dipole models. We treat the effective temperature, the dipole
  strength $B_d$, and the offset $a_z$ (in units of stellar radii) as
  free parameters. The top and bottom spectra show the inner and outer
  pair of the split components, respectively. }
\label{fig:fit}
\end{figure}

Some of the Gemini spectra have both the inner and outer pair of lines
visible; we fit only the strongest pair of absorption features in
those cases. Figure \ref{fig:fit} shows our model fits to two of the
G183$-$35 spectra. The top spectrum shows the inner pair of lines,
which indicate a dipole field strength of $B_d=8.6$ MG, whereas the
bottom spectrum shows the outer pair of lines, which indicate a field
strength of $B_d=$10.9 MG.

\begin{figure}
\vspace{-0.7in}
\includegraphics[width=3.4in]{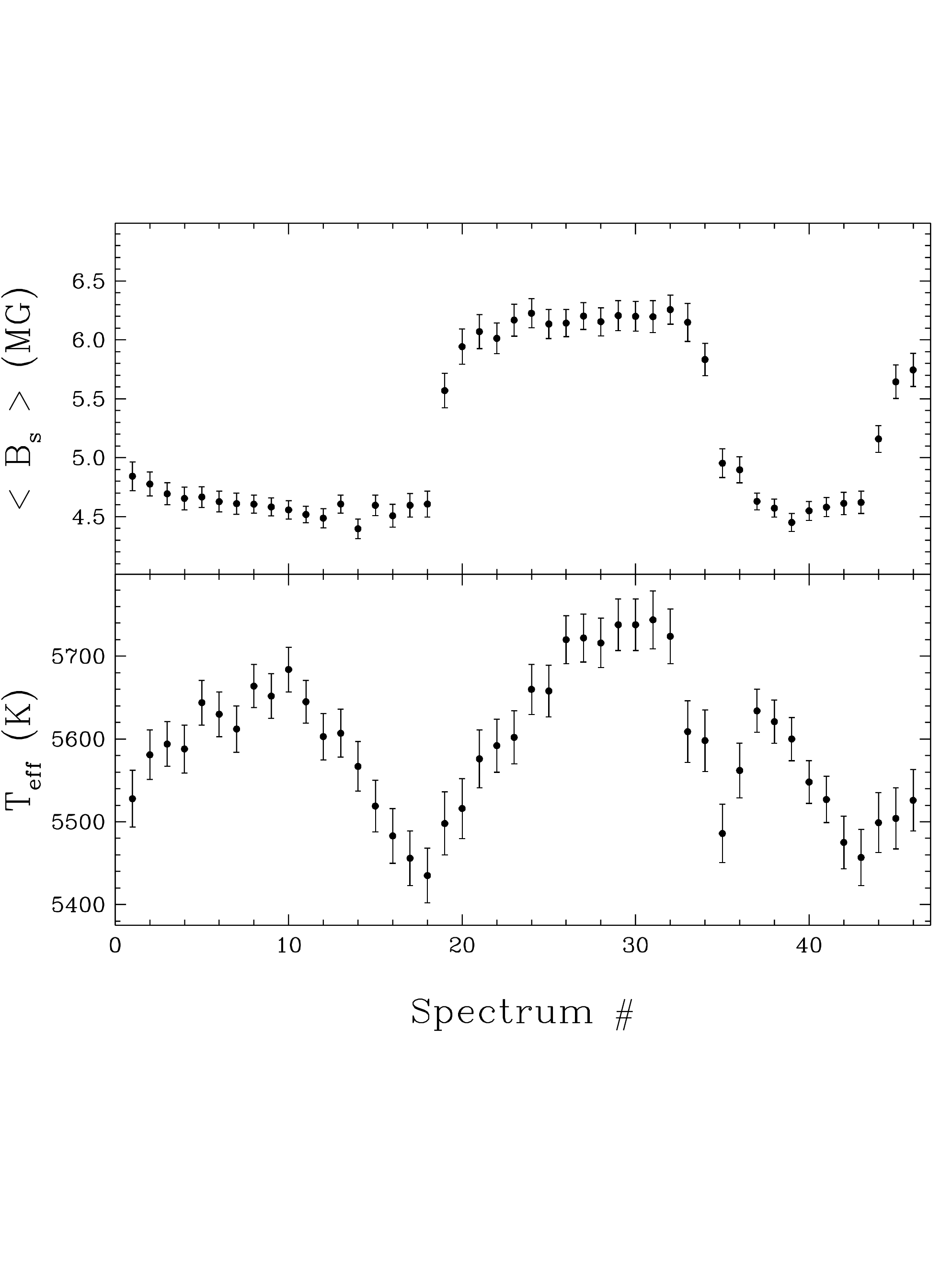}
\vspace{-0.8in}
\caption{The average field strength over the visible stellar disk (top
  panel) and effective temperature (bottom panel) of the best-fit
  magnetic white dwarf model for each Gemini spectrum. Note that the
  first 32 spectra were obtained on the first night, the next 4 on the
  second night, and the last 10 were obtained on the last night of
  Gemini observations.}
\label{fig:az}
\end{figure}

Figure \ref{fig:az} shows the evolution of the mean field modulus
(i.e., the average of the field strength over the visible stellar
disk) and the best-fit temperature as a function of time. G183$-$35
switches between a low mean field modulus of $\approx$4.6 MG and a
high mean field modulus of $\approx$6.2 MG over several hours. In
addition, we find variations in the effective temperature of the
best-fit model, which is a manifestation of the changes in the central
H$\alpha$ component. Of course, a time-dependent dipolar field
strength is not realistic. Instead, the variations observed here are
most likely due to rotation, with the magnetic axis offset with
respect to the rotation axis, a model known as the oblique rotator
\citep{stibbs50,monaghan73}. Because these two axes are not aligned, the
observer sees a different magnetic field distribution across the
stellar surface due to rotation.

There are several previously known examples of magnetic white dwarfs
that display rapid evolution in their spectral line profiles.  PG
1031+234 is one such system where \citet{latter87} observed
significant changes in the spectrum of this object over the spin
period of 3 h 24 min. They find two spectral features with field
values $\sim$200 MG between rotational phases of 0.5 to 0.1, which
then diffuse out and disappear at phase 0.15 as very high field zones
appear with a field strength of 1000 MG and remain visible for phases
of 0.15 to 0.5. \citet{latter87} conclude that their spectroscopic
data is best explained by a field pattern with a slightly offset 500
MG global component with a localized magnetic spot with a central
field of nearly 1000 MG.
 
EUVE J0317$-$85.5 is another rapidly changing system with a spin period of 12 min \citep{barstow95}. 
\citet{vennes03} display far ultraviolet time-series spectroscopy
of EUVE J0317$-$85.5 in their Figure 6. The spectra show the 1s0-2p-1 component of Ly $\alpha$ near 1300 \AA\ between
the phases of 0.3 and 0.7, but this line rapidly shifts to 1340 \AA\ and remains at that wavelength between the
phases 0.7 and 0.3. An overlap of  high-field and low-field features is only apparent at phases 0.3 and 0.7.
\citet{vennes03} conclude that a high-field ($B \geq 425$ MG) magnetic spot with underlying lower field ($B \leq 185$ MG)
surface would explain the variability in this system, including a rapid transition from low-field to high-field line spectra.

WD 1953$-$011 has a rotation period of 1.448 d, and its H$\alpha$ line profile is also best-explained by a two-component magnetic field that includes a weak (180-230 kG), large-scale component, and a strong
(520 kG), localized component, i.e. a spot \citep{maxted00,valyavin08}. The large scale component is almost always visible through the
narrow splitting of the central H$\alpha$ line, and the spot is visible only at rotational phases of 0.25-0.7 through two broad features at
6554 and 6576 \AA. Hence, the combined spectrum of WD 1953-011 would also display an H$\alpha$ line split into five components.
\citet{valyavin08} find evidence for rotational variability of the projected effective size of the magnetic spot ranging
from 0 to 12\% of the disk. Interestingly, the appearance/disappearance of the strong field component is very similar to the variability
seen in G183$-$35. However, unlike in G183$-$35, the radial velocity of the central H$\alpha$ component in WD 1953-011
is constant to within a few km s$^{-1}$ \citep{maxted00}. The similarities between the spectral evolution of G183$-$35 and the
three other examples presented here strongly favor rotational modulation as the source of variability in G183$-$35. 

\subsection{Comparison with Ap/Bp stars}

About 10\% of A and B type main-sequence stars host detectable magnetic fields. The majority of these stars are chemically
peculiar, and therefore classified as Ap/Bp stars. These stars also show variations in their magnetic field strengths, spectral
line profiles, and luminosities on timescales related to their rotation periods \citep[][and references therein]{bailey15a}. 
Many of these stars show abundance variations over the stellar surface which give rise to changes in their spectral line
profiles. Analyzing spectropolarimetry of three such stars,
\citet{kochukhov17} and \citet{kochukhov19} find distortions in the spectral line profiles of several metal lines that indicate
large-scale, high-contrast abundance patterns over the stellar surface. They also detect significant changes in the magnetic
field strength and topology, switching between a low-field and a high-field structure on the rotation period. 

\citet{bohlender10} detected significant changes in the He line profiles of the Bp star a Centauri (HR 5378), and found
that the He abundance geometry is consistent with a single spot model where one hemisphere of the star has an enhanced He
abundance while the other hemisphere is He deficient. Similarly, \citet{bailey12} found H$\alpha$ line profile variations at
different rotation phases in HD 133880. The line core shows excess absorption or emission compared to the average profile,
which could be interpreted as radial velocity variations. Interestingly, the variations are only seen in the core of the
H$\alpha$ line and they closely mimic the variations observed in Fe lines. \citet{bailey12} conclude that HD 133880 may
be similar to a Centauri and it may also suffer from abundance anomalies between the different sides of the star. 

\citet{bailey15b} report the detection of radial velocity variations of up to 35 km s$^{-1}$ in the magnetic Ap star HD 94660.
They emphasize that many Ap/Bp stars show variations due to shifts in the centre-of-gravity of the line profile due to the inhomogeneous surface distribution (e.g. spots), but these shifts are always smaller than the width of the line, and they also follow the rotation
of the star. In the case of HD 94660 the rotation period is $\approx$2800 d, whereas the radial velocities vary with
a $\sim$840 d period. Hence, the variations seen in this star is likely due to binarity.

Like most Ap/Bp stars, G183$-$35 shows variations in its magnetic field strength, spectral line profiles, and luminosity over
a period of about 4 h. Hence, these variations are almost certainly due to the spin of the white dwarf.
The distortions in the central H$\alpha$ component, which could be interpreted
as radial velocity variations, can be explained by an inhomogeneous surface H distribution. He becomes invisible below about 11,000
K in white dwarf atmospheres. Hence, it is possible that G183$-$35 has a mixed H/He atmosphere with patchy H, and
abundance variations across the stellar disk could lead to the observed distortions in the H$\alpha$ line. 

It is intriguing that the majority of the magnetic DAs analyzed by \citet{rolland15} are all found in the same temperature range, between
5000 and 6000 K, in the so called non-DA gap \citep{bergeron97}. It is also suspicious that the H$\alpha$ line profiles for the majority
of these magnetic white dwarfs require dilution by a DC companion. There is a simpler explanation for the unusual line profiles;
a chemically inhomogeneous mixed H/He atmosphere.
\citet{pereira05} found quasi-periodic variations in the strengths of the H and He lines over a period of $\sim$3.5 h in the DAB
white dwarf GD 323. They found that a model with an inhomogeneous surface composition, resulting from the dilution of a thin hydrogen atmosphere with the underlying helium convection zone, best matches the observations. Hence, it is possible that G183$-$35 and
the other unresolved binary candidates presented in \citet{rolland15} have inhomogeneous surface composition with patchy H.
Such a scenario would explain the observed variations in the line profiles and the discrepancy between the photometric and
spectroscopic temperature measurements. In this scenario, only a fraction of the star would contribute to the H lines, and
the temperature variations seen in Figure \ref{fig:az} could be a manifestation of that. Follow-up observations of the
other binary white dwarf candidates in \citet{rolland15} can test this scenario.

\section{Conclusions}

We have presented time-series spectroscopy and photometry of the magnetic white dwarf G183$-$35. Even though the average
spectrum shows an H$\alpha$ line split into 5 components, most spectra show only the inner or the outer pair of lines.
The radial velocities of the central component, equivalent widths of the inner and outer pair of lines, and the photometry all show
variations on a period of $\sim$4 h. Orbital motion cannot explain the amplitude of the radial velocity variations and also
the appearance and disappearance of the different sets of lines. On the other hand, rotation of a magnetic white dwarf with a
chemically inhomogeneous surface, much like in Ap/Bp stars, can explain both spectroscopic and photometric variations seen
in this star.  Spectropolarimetry of G183$-$35 would help in understanding this object further by constraining its field topology.

\section*{Acknowledgements}

We thank Jim Fuller for useful discussions, and the referee, Stefano Bagnulo, for helpful suggestions.
We gratefully acknowledge the support of the NSF under grant AST-1906379.
This work is supported in part by the NSERC Canada and by the Fund FRQ-NT (Qu\'ebec).
Based on observations obtained at the Gemini Observatory, which is operated by the Association of Universities for Research in Astronomy, Inc., under a cooperative agreement with the NSF on behalf of the Gemini partnership: the National Science Foundation (United States), National Research Council (Canada), CONICYT (Chile), Ministerio de Ciencia, Tecnolog\'{i}a e Innovaci\'{o}n Productiva (Argentina), Minist\'{e}rio da Ci\^{e}ncia, Tecnologia e Inova\c{c}\~{a}o (Brazil), and Korea Astronomy and Space Science Institute (Republic of Korea).

\bibliography{master}

\bsp
\label{lastpage}

\end{document}